\documentclass{article}

\usepackage[final,nonatbib]{neurips_2018}

\usepackage[utf8]{inputenc} 
\usepackage[T1]{fontenc}    
\usepackage{hyperref}       
\usepackage{url}            
\usepackage{booktabs}       
\usepackage{amsfonts}       
\usepackage{nicefrac}       
\usepackage{microtype}      
 \usepackage[pdftex]{graphicx}
\graphicspath{ {./images/} }

\usepackage[numbers]{natbib}

\title{Keyword Spotter Model for Crop Pest and Disease Monitoring from Community Radio Data}

\author{
  Benjamin Akera\\
  Makerere University\\
  \texttt{akeraben@gmail.com} \\
     \And
   Joyce Nakatumba-Nabende \\
  Makerere University\\
  \texttt{jnakatumba@cis.mak.ac.ug} \\
   \And
    Jonathan Mukiibi \\
  Makerere University\\
  \texttt{jonmuk7@gmail.com} \\
    \And
  Ali Hussein \\
  Ronin Institute \\
  \texttt{ali.hussein@ronininstitute.org} \\
   \And
  Nathan Baleeta \\
  Makerere University \\
  \texttt{nbaleeta@gmail.com} \\
   \And
  Daniel Ssendiwala \\
  Makerere University \\
  \texttt{ssendiwaladaniel@gmail.com} \\
    \And
  Samiiha Nalwooga \\
  Makerere University \\
  \texttt{nsamiiha@gmail.com} \\
}

\begin{document}

\maketitle

\begin{abstract}

In societies with well developed internet infrastructure, social media is the leading medium of communication for various social issues especially for breaking news situations. In rural Uganda however, public community radio is still a dominant means for news dissemination.
Community radio gives audience to the general public especially to individuals living in rural areas, and thus plays an important role in giving a voice to those living in the broadcast area. It is an avenue for participatory communication and a tool relevant in both economic and social development.This is supported by the rise to ubiquity of mobile phones providing access to phone-in or text-in talk shows. In this paper, we describe an approach to analysing the readily available community radio data with machine learning-based speech keyword spotting techniques. We identify the keywords of interest related to agriculture and build models to automatically identify these keywords from audio streams. Our contribution through these techniques is a cost-efficient and effective way to monitor food security concerns particularly in rural areas. Through keyword spotting and radio talk show analysis, issues such as crop diseases, pests, drought and famine can be captured and fed into an early warning system for stakeholders and policy makers.

\end{abstract}

\section{Introduction}
Ensuring a functional and near real-time system of surveillance for crop diseases and pests is of critical importance to sustaining the livelihoods of smallholder farmers in sub-Saharan Africa \citep{mutembesa2018crowdsourcing}. Disease and pest surveillance systems have to be put in place to provide early warning to the farmers and the relevant agricultural research bodies. Usually, when a crop disease or pest is reported in a given area, experts from the respective research institutes take time to reach the reported location to carry out investigations. This usually involves inspection of the crops at specific intervals (of about 10 km) along the more accessible main roads, covering only small proportions of the areas of interest in major districts \citep{mutembesa2018crowdsourcing}. Because the surveillance teams have to work within limited budgets, the surveys and the results from the surveys may be delayed or fewer regions may be sampled in a particular year. As the health experts provide an annual snapshot of the health of crops across the country they are limited in their ability to provide real-time actionable surveillance data.
In many cases, the farmers never get to know the disease that has attacked their crops for weeks or even months.

In many areas in Uganda, the vast majority of the affected people will use social media to communicate their    concerns in their local communities. This social media is not Facebook or Twitter, its the local community radio stations existing in almost each village in sub-Saharan Africa \citep{saeb2017very}. These rural radio stations give farmers an opportunity to interact with each other and also with the relevant agricultural authorities such as extension workers and agricultural experts. This can  usually be through a number of formats like phone-in programs, live talk shows \citep{nakabugu2001role}. Specifically for some of the radio stations, they have targeted agricultural talk shows that can host an expert from an agricultural research institute who can aim at providing specific information for example: about crop disease and pest management.

Keyword Spotting systems (KWS) is a classification task that aims at detection and retrieving of a series of words from a database of audio streams. The advantage of using a KWS is that unlike full automatic speech recognition systems, they can be developed without sufficient labelled data. This is common especially in low resourced languages \cite{menon2019}.  In this paper, we discuss an implementation of a Keyword Spotting model that we use to mine local community radio content using specific keywords for a low resourced language in Uganda. We evaluate our approach on the Luganda language which is a low-resource language that is currently spoken and used in many of the Agricultural communities in Uganda.

\section{Related Work}
Previous work investigating crop disease surveillance utilizes different approaches. Some approaches have focused on setting up a crop disease surveillance network that relies on the use of mobile phones \cite{mutembesa2018crowdsourcing,mutembesa2019mobile} while others use satellite imagery \cite{zhang2014monitoring}. The disease detection aspect of the surveillance module uses computer vision and machine learning to detect plant diseases based on leaf imaging \cite{aduwo2010automated,mwebaze2016machine}. Leaf-based  approaches however rely on the use of imaging device for low-resource approaches utilizing smartphones \cite{mutembesa2018crowdsourcing,mutembesa2019mobile,quinn2011modeling,Quinn2013}. This may be limited in areas with no or low smartphone adoption.

Keyword spotting for low resource languages has been implemented before \cite{menon2017radio,saeb2017very}. The  approaches include used include CNNs, siamese CNNs \cite{bromley1994signature} and autoencoders. Other models are designed for low computational resources such as \cite{deepresidual,dilated_convs} due to the popularity of using keyword spotting to identify commands in smartphones, battery concerns from CPU requirements comes into play. Low resource languages pose a problem as models that consume a lot of data in training fail to converge due to low volume of text corpora and speech recording. 

Luganda is an almost zero-resource Bantu language, spoken in the central region of Uganda. Work on Luganda remains small compared to larger languages such as Kiwahili, Zulu and Hausa. Research on Luganda exists in machine translation \cite{luganda_text_to_speech}, keyword spotting \cite{menon2017radio,saeb2017very}. Prior work on Luganda keyword spotting and radio monitoring of Luganda community radio has been initiated by \cite{menon2017radio,saeb2017very} to generate insights concerning humanitarian aid and development. While radio content is publicly available and accessible with seemingly no data/privacy restrictions, there have been few interventions seeking to mine this data for surveillance purposes particularly for crop pests and diseases.

\section{Methodology}
\subsection{Building the Keyword Corpus}
The primary source of keywords in this study are radio recordings captured from radio stations spread across Uganda. We selected 55 radio stations which are commonly listened to in the central region. For each radio station, a google search was done to find out whether it had an online radio station and whether it had its radio schedule available online. From the initial list of radio stations, 19 of them had online streams and of these at least 14 radios broadcast in Luganda. We identified radio schedules for 10 of these radio stations from the radio station websites and also by manually listening in to the stations. The purpose was to identify the time when there are talk shows particularly the agricultural ones. It was observed that for most talk shows topics of discussions are picked depending on the audience demand, the trending topic in the society or country, sponsors/advertisers or specific campaigns though still there are weekly talk shows which are focused on agriculture.
 The purpose was to identify the time when the talk shows were hired particularly the agricultural ones.
 
 We collected data from online radio stations and these were recorded as 5-minute audio clips which were stored in a shared Dropbox folder. The team also identified 2 radio stations which avail their radio content online for the past 7 days as 1-hour recordings. The websites were scrapped and the audio recordings were sorted depending on whether the 1 hour was a talk show or not. After which, the talk shows were trimmed into 5 minute audio clips. 
 The 5-minute audio clips were then played back and carefully listened to by a team of three volunteers with the purpose of identifying and extracting the commonly used agricultural terms that would be fed into the keyword spotting model. 
 
 To compliment the keywords captured from radio talkshows, we also scrapped an online local newspaper in Luganda commonly known as Bukedde\footnote{\url{https://www.bukedde.co.ug}}. 
 One advantage of using online articles as a source of keywords is that there are different ways in which the same crop disease or pest is mentioned and the spelling of such words can be  to is captured specifically for Luganda. 
 The keywords were then grouped into crops, diseases, fertilizers, herbicides and general keywords. Translations were also added in two languages that is Luganda and English as well as an alternative keywords in form of a stem and this was in case the stem alone was a unique keyword.


Both the keyword sets from the local radio and online sources were then aggregated into one keyword corpus of 193 keywords. Table \ref{keywordref} shows an example of keywords extracted from a radio talkshow on the Fall Army Worm pest which is affecting maize.

\begin{table}[!htbp] 
  \caption{Aggregated Keywords from a sample radio talkshow on the fall army worm pest.}
  \centering
  \begin{tabular}{lll}
    \toprule
    \multicolumn{2}{c}{Sample Keywords}                   \\
    \cmidrule{1-2}
    Keyword     & Description     & Frequency (per recording)\\
    \midrule
    Akasanyi & Luganda term meaning \textit{worms}  & 18 times     \\
    Obutunda     & Luganda term meaning \textit{passion fruits} & 22 times      \\
    Kasooli     & Luganda word meaning \textit{maize}      &  35 times  \\
    \bottomrule
  \end{tabular}
  \label{keywordref}
\end{table}

\subsection{Speech Keyword Dataset}

Audio data collection was performed by crowdsourcing speech utterances of the different words from the keyword list. The use of studio captured samples seemed unrealistic, to mimic real-world settings the data was collected in a natural setting with noisy environments, poor quality recording equipment, and people talking in a natural, chatty way. Rather than using high quality microphones, and in a formal setting. This was ensured through the audio data collection tool derived from \cite{open_speech} where the person speaking out the words can do it using their phone or laptop wherever they are. In this study, we collected data from over 35 users who recorded the keywords in Luganda and English.

An important goal here was to record sufficient data to train the model but low enough to allow for low-resource training. We ensured that we averaged 10 utterances per keyword. Keyword spotting models are much more useful if they are speaker independent, since the process of personalizing a model to an individual requires an intrusive user interface experience. With this in mind, the recording process had to be quick and easy to use, to reduce the number of people who would fail to complete it. The collected keyword audio data was encoded in \textit{ogg vorbis} format.

\subsection{Data Preprocessing}

In order to perform speech processing, our first step is to convert the recorded ogg keyword files to wav files. As ogg is a lossy encoding format, we used ffmpeg \cite{ffmpeg} to decode the ogg vorbis files into wav audio files. The files were then transformed into 1d vectors using librosa \cite{mcfee2015librosa}. Then for an audio signal $w_{s_t}$ with a sampling rate $s$ and a length $t$. We use  \cite{mcfee2015librosa} resampling feature shown: 
\begin{equation}
f_{re}:w_{s_t}\rightarrow w_{8kHz_t} 
\end{equation}
to normalize the sampling rate of all the samples to 8kHz. 

\section{Model Architecture and Design}
In this study, We use a 1-dimensional Convolutional Neural Network (CNN), a siamese Convolutional Neural Network (CNN), and time delay networks to implement keyword spotters.

\subsection{1-d CNN}

The 1-d CNN takes in as input the processed raw audio data. The input to the model is an array representing the audio waveform ($X$). The network is designed to learn the set of parameters ($\theta$) to map the input to a prediction ($T$) according to a hierarchical feature extraction given by equation 2. 

\begin{equation}
    T = F(X|\Theta ) = f_L(...f_2(f_1(X|\theta_1)|\theta_2)|\theta_L)
\end{equation}

where $L$ is the number of hidden layers in the network.

The final architecture we created is  a 14 layer deep neural network with five 1D convolutional layers with each with intermediate pooling layers. A dropout of 0.5 was also applied after the two successive dense layers. The final layer is a softmax activation function to map to only 10 target keywords selected randomly from the Luganda/English corpus. 
The model was trained using batch gradient descent with the Adam Optimizer and a learning rate of 0.001. 

\section{Results}
As a baseline experiment we trained a baseline 5-layer densely connected network and our 1d convolutional model. We trained the model with early stopping at no improvement in validation loss for 10 epochs, model stopped at 30 epochs. We trained our model on a total of 18426 samples and validated it on 4607 and tested on 5759 samples. This model was trained on a K80 GPU on Google Colab. The evolution of loss and accuracy are shown in Figure 1. Results are shown in Table \ref{results}.

\begin{table*}[!h]
	\caption{Keyword Classification Prediction Results}
	\label{tab:freq}
	\begin{tabular}{ccccl}
		\toprule
		Model &Accuracy &Average Precision &Average Recall&F1 Score\\
		\midrule
		Dense Baseline & 0.62 & 0.62 & 0.60 & 0.60\\
		1d-conv & 0.93 & 0.93 & 0.93 & 0.93\\
		\bottomrule
	\end{tabular}
	  \label{results}
\end{table*}

The results show that the 1D convnet performs better than the Dense baseline model used.  We see an improvement in Accuracy from 0.62 to 0.93 and similar improvement in the Average Precision and Averagae recall. The 1D convnet gives us better precision by reducing the false positive rate observed by the Dense Baseline model.

\section{Conclusion and Future Direction}

Using 1D convolutions for low-resource keyword spotting shows promising results. We aim to explore the usability of this work with other Bantu languages, specifically those geographically close to Luganda for example Runyankore, Kinyarwanda and Tooro as well as other linguistically further but geographically close languages such as Luo and Karamojong. We are also interested in the possibility of using transfer learning in keyword spotting in different dialects of the same language as in Acholi, Lango and Kumam for Luo and different but linguistically related languages such as Runyankore.

\bibliographystyle{plain}

\begin{thebibliography}{99} 
\bibitem[1]{mutembesa2018crowdsourcing}  Mutembesa, D. and Omongo, C. \& Mwebaze, E.\ (2018) Crowdsourcing real-time viral disease and pest information: A case of nation-wide cassava disease surveillance in a developing country. {\itshape Sixth AAAI Conference on Human Computation and Crowdsourcing} 
    
\bibitem[2]{mutembesa2019mobile} Mutembesa, D., Mwebaze, E., Nsumba, S., Omongo, C. \& Mutaasa, H.\ (2019) Mobile community sensing with smallholder farmers in a developing nation; A scaled pilot for crop health monitoring., {\itshape arXiv preprint arXiv:1908.07047}

\bibitem[3]{menon2017radio} Menon, R., Saeb, A., Cameron, H., Kibira, W., Quinn, J., \& Niesler, Thomas\ (2017) Radio-browsing for developmental monitoring in Uganda. In G.\ Tesauro, D.S.\ Touretzky and
T.K.\ Leen (eds.), {\itshape IEEE International Conference on Acoustics, Speech and Signal Processing (ICASSP)}, pp.\ 5795--5799. IEEE
  
\bibitem[4]{saeb2017very} Saeb, A., Menon, R., Cameron, H.,  Kibira, W., Quinn, J. \& Niesler, Thomas.\ (2017) Very Low Resource Radio Browsing for Agile Developmental and Humanitarian Monitoring., {\itshape INTERSPEECH}, pp.\ 2118--2122. 


\bibitem[5]{nakabugu2001role} Nakabugu, S.B.\ (2001) The role of rural radio in agricultural and rural development translating agricultural research information into messages for farm audiences, {\itshape First International Workshop on Farm Radio Broadcasting}, pp.\ 19--22. 


 \bibitem[6]{mwebaze2016machine} Mwebaze, E.\ \& Owomugisha,  G.\ (2016) Machine learning for plant disease incidence and severity measurements from leaf images., {\itshape 15th IEEE International Conference on Machine Learning and Applications (ICMLA)}, pp.\ 158--163. IEEE.

 \bibitem[7]{Quinn2013} Quinn, J. (2013) Computational Techniques for Crop Disease Monitoring in the Developing World., {\itshape Advances in Intelligent Data Analysis {XII}}, pp.\ 13--18. Springer Berlin Heidelberg.


\bibitem[8]{quinn2011modeling} Quinn, J.A., Leyton-Brown, K. \& Mwebaze, E.\ (2011) Modeling and monitoring crop disease in developing countries, {\itshape Twenty-Fifth AAAI Conference on Artificial Intelligence}

\bibitem[9]{deepresidual} Tang, R.\ \& Lin,  J..\ (2018) Deep Residual Learning for Small-Footprint Keyword Spotting., {\itshape IEEE International Conference on Acoustics, Speech and Signal Processing (ICASSP)}, pp.\ 609--616. Cambridge, MA: MIT Press.


\bibitem[10]{dilated_convs} Coucke, A., Chlieh, M., Gisselbrecht, T., Leroy, D., Poumeyrol, M. \& Lavril, T.\ (2019) Efficient Keyword Spotting Using Dilated Convolutions and Gating, {\itshape IEEE International Conference on Acoustics, Speech and Signal Processing (ICASSP)}, pp.\ 609--616. Cambridge, MA: MIT Press.

\bibitem[11]{luganda_text_to_speech} Nandutu, I.\ (2016) Luganda Text-to-Speech Machine., {\itshape School of Computing and Engineering, Uganda Technology and Management University}.

\bibitem[12]{mcfee2015librosa} McFee, B., Raffel, C., Liang, D., Ellis, D.P.W., McVicar, M., Battenberg, E. \& Nieto, O.\ (2015) Librosa: Audio and music signal analysis in python, {\itshape Proceedings of the 14th python in science conference}

\bibitem[13]{ffmpeg} FFmpeg, {\itshape {\url{http://www.ffmpeg.org/}}}.


\bibitem[14]{open_speech} Warden, P.\ (2017) Open Speech Recording, {\itshape {\url{https://github.com/petewarden/open-speech-recording}} }
   
   
\bibitem[15]{menon2019} Menon, R., Kamper, H.,  van der Westhuizen,  E.,  Quinn, J. \& Niesler, T.\ (2019) Feature exploration for almost zero-resource ASR-free keyword spotting using a multilingual bottleneck extractor and correspondence autoencoders, {\itshape INTERSPEECH}


\bibitem[16]{aduwo2010automated} Aduwo, J. R.,  Mwebaze, E. \& Quinn, J. A.\ (2010) Automated Vision-Based Diagnosis of Cassava Mosaic Disease., {\itshape Industrial Conference on Data Mining-Workshops}, pp.\ 114--122. 

\bibitem[17]{zhang2014monitoring} Zhang, J., Pu, R., Yuan, L., Wang, J., Huang, W., \&  Yang, G.\ (2014) Monitoring powdery mildew of winter wheat by using moderate resolution multi-temporal satellite imagery., {\itshape PloS one}, 9(4),pp.\ e93107. Public Library of Science.
   
 
 \bibitem[18]{bromley1994signature} Bromley, J., Guyon, I., LeCun, Y. and S{\"a}ckinger, E., Shah, R.\ (1994) Signature verification using a "siamese" time delay neural network., {\itshape Advances in neural information processing systems}, 9(4),pp.\ 737--744.
   
\end{thebibliography}

\end{document}